 \documentclass[conference]{IEEEtran}

\usepackage{graphicx}
\usepackage[para]{threeparttable}
\usepackage{subfig}
\usepackage{array}
\usepackage{mathtext}

\usepackage{amsmath}
\usepackage{algorithmic}
\usepackage{algorithm}
\usepackage{url}
\usepackage{epstopdf}
\usepackage{cite}
\usepackage{setspace}
\usepackage{graphicx,lipsum}
\usepackage{caption}
\usepackage{tikz}

\newcolumntype{P}[1]{>{\centering\arraybackslash}p{#1}} 
\newcolumntype{M}[1]{>{\centering\arraybackslash}m{#1}} 

\usepackage{xcolor}
\newcommand{\ws}[1]{\textcolor{blue}{#1}}
\renewcommand{\ws}[1]{#1}
\newcommand{\wx}[1]{\textcolor{red}{#1}}
\renewcommand{\wx}[1]{#1}

\begin{document}
\bibliographystyle{ieeetr}
\title{Statistical Relationship between Interference Estimates and Network Capacity}
\author{\IEEEauthorblockN{Srikant Manas Kala$^\dag$, Winston K.G. Seah$^\Phi$, Vanlin Sathya$^*$, Betty Lala$^\P$}
\IEEEauthorblockA{$^\dag$Indian Institute of Technology Hyderabad, India. $^\Phi$Victoria University of Wellington, New Zealand.\\ $^*$University of Chicago, Illinois, USA. $^\P$Kyushu University, Fukuoka, Japan.\\
Email: cs12m1012@iith.ac.in, winston.seah@ecs.vuw.ac.nz, vanlin@uchicago.edu, 3es18314s@s.kyushu-u.ac.jp}}

\maketitle
\begin{abstract}

Interference is a major impediment to the performance of a wireless network as it has a significant adverse impact on Network Capacity. \wx{There has been a gradual and consistent densification of WiFi networks due to Overlapping Basic Service Set (OBSS) deployments. With the upcoming 802.11ax standards, dense and ultra-dense deployments will become the norm and the detrimental impact of Interference on Capacity will only exacerbate.} However, the precise nature of the association between Interference and Network Capacity remains to be investigated, a gap we bridge in this work. We employ linear and polynomial regression to find answers to several unexplored questions concerning the \textit{Capacity Interference Relationship} (CIR). We devise an algorithm to select regression models that best explain this relationship by considering a variety of factors including \textit{outlier threshold}. We ascertain the statistical significance of their association, and also determine the explainability of variation in Network Capacity when Interference is varied, and vice versa. While the relationship is generally believed to be non-linear, we demonstrate that scenarios exist where a strong linear correlation exists between the two. We also investigate the impact of WMN topology on this relationship by considering four carefully designed Wireless Mesh Network (WMN) topologies in the experiments. To quantify endemic Interference, we consider four popular Theoretical Interference Estimation Metrics (TIEMs) \emph{viz.,} \textit{Total Interference Degree} (TID), \textit{Channel Distribution Across Links Cost} (CDAL$_{cost}$), \textit{Cumulative X-Link-Set Weight} (CXLS$_{wt}$), and \textit{Channel Assignment Link-weight Metric} (CALM). To ensure a sound regression analysis, we consider \ws{a large} 
set of 100 Channel Assignment (CA) schemes, a majority of which are generated through a Generic Interference-aware CA Generator proposed in this work.  Finally, we test the TIEMs in terms of their reliability and the ability to model Interference. We carry out the experiments on IEEE 802.11g/n WMNs simulated in ns-3.  
\end{abstract}
\section{Introduction}

\wx{There has been a phenomenal 17-fold increase in the global mobile data between 2012 and 2017, and a 71\% rise in 2017 alone. There were speculations that with the increase in penetration of 4G/LTE and 5G networks, WiFi will no longer be the prime driver of data demand. The numbers are in, and despite the much higher speeds and relaxed data caps that 5G networks will boast of, WiFi is here to stay for the foreseeable future. Over half of all data traffic on 4G was offloaded on to WiFi in 2017, which is expected to rise to 59\% by 2022. 5G deployments will be even more dependent on WiFi, as over 71\% data offload is expected by 2022.}

\wx{The global mobile data traffic is estimated to rise to a staggering 77 exabytes/month by the end of 2022. With a simultaneous rise in Wi-Fi offloading by the telecom networks, the WiFi networks have become increasingly dense, and even ultra-dense with inter-AP (Access Point) distance now less than 10m in the urban centers of countries such as Japan. Such dense and ultra-dense deployments often have Overlapping Basic Service Set (OBSS) owing to a higher density of APs whose coverage area overlaps. Although the OBSS deployments facilitate better spatial reuse, higher order modulations, and better signal strength, the AP coverage overlap creates plethora of performance bottlenecks, the most significant of which is co-channel Interference.}

\wx{A study group has been tasked with the release of the IEEE802.11ax standard for dense and ultra-dense Wireless Mesh Networks (WMNs) which will cater to the increased data demand by ensuring enhanced system performance in terms of network throughput and spectrum efficiency. Dense deployments comprising of multiple OBSS will enable the 802.11ax amendment to offer High Efficiency WLAN (HEW) services to end-users.}

\wx{However, due to the dense OBSS deployment, several challenges posed by interference, which have been successfully addressed in conventional 802.11 networks, will require to be addressed anew in the 802.11ax supported HEWs. For example, the problem of hidden nodes will acquire new dimensions as the RTS-CTS mechanism in its current form is incapable of addressing the challenges of co-channel interference and hidden-node problems in multi-AP OBSS deployments. Likewise, other problems cause by endemic interference that adversely impact network performance will need to be revisited in dense and ultra dense networks, \emph{viz.}, AP-AP Interference, Interference amplification effect, exposed node problem,  Interference deadlock, deadlock and link suppression \emph{etc.} }

\wx{Although, the overarching challenge of OBSS management has been given due consideration in all working groups commissioned by IEEE since the 802.11aa task group, the detrimental impact of interference in dense OBSS deployments requires special emphasis. Several recent works have begun addressing the interference related challenges in dense and ultra-dense WiFi deployments, and we briefly review a few state-of-the-art works. Authors in \cite{dense1} investigate the effect of inter-AP distance on network performance in dense OBSS scenarios, and state that increasing the density of APs does not necessarily translate into improvement in network capacity. They recommend designing efficient load balancing and channel assignment mechanisms that specifically cater to 802.11ax HEWs. The work also highlights several interference related challenges in context of the upcoming dense OBSS deployments, but an analysis of the  relationship between network capacity and Interference is lacking. 

In \cite{dense2}, authors propose the use of dynamic sensitivity control for adaptive Clear Channel Assessment, among other solutions, to maximize network capacity in ultra-dense WiFi networks. In a recent prominent work on 802.11ax \cite{dense4}, interference is stated to be the primary challenge to dense networks. Likewise, expected challenges to 802.11ax HEWs from internal interference, WiFi-LTE coexistence, and interference in mobile/vehicular environments are discussed \cite{dense3}.}

\wx{However, these works fall short of investigating the Capacity Interference Relationship (CIR) in conventional WiFi deployments or current dense and ultra-dense networks.
The same goes for the earlier research literature on Interference mitigation to enhance network performance of conventional WiFi deployments. Almost invariably, the objective of interference alleviation strategies was to enhance the \textit{Network Capacity}, while secondary objectives included a reduction in packet loss and end-to-end latency. The determination of the maximal achievable throughput in a wireless network, under the adverse impact of endemic Interference and mechanisms employed to alleviate it, is a non-trivial NP-hard problem \cite{npt}. So, research studies have attempted to explore an approximate relationship between Interference and Network Capacity and have revealed an inverse relationship between the two \cite{2Gupta, Capacity}. Some studies have even suggested an upper-bound on maximal achievable throughput \cite{Impact}, but they require several vital inputs about the network beforehand, such as the network layout and the location of static nodes, expected traffic load \emph{etc.} It is well established that the Network Capacity and the intensity of Interference prevalent in a wireless network are inversely related, \emph{i.e.,} to improve network throughput the transmission conflicts must be reined in. However, an exact quantification of this inverse relationship that is not specific to the network layout, or limited to a theoretical upper-bound, is lacking.}

To the best of our knowledge, a statistical analysis of the relationship between Capacity and Interference based on empirical observations is lacking. \wx{We contend that with the densification of WiFi deployments and the emergence of 802.11ax standard, such an analysis is not only relevant, but of great necessity. In this work we bridge this gap by conducting CIR analysis in four conventional (802.11g/n) Wireless Mesh Networks (WMN) through the statistical tools of linear and polynomial regression. The findings of this work are counter-intuitive and hold great relevance to practical applications in dense and ultra-dense WiFi deployments.}

To measure the impact and intensity of prevalent interference, we make use of four commonly used Theoretical Interference Estimation Metrics (TIEMs). TIEMs offer a theoretical measure of the Interference prevalent in a wireless network, and its adverse impact on network performance. Several TIEMs have been devised in earlier studies to offer a measure of Interference prevalent in WMNs. The most commonly used TIEM is the Total Interference Degree \cite{TID1}, which is a measure of all \textit{potential} link-conflicts present in the conflict graph of the wireless network. 
Over the years, countless \textit{Interference-aware} Channel Assignment (CA) schemes have been proposed based on this guiding principle to maximize the network throughput \cite{22Ramachandran,24Aizaz, Manas2}.
The \textit{Channel Distribution Across Links} (CDAL) approach offers an Interference metric called CDAL$_{cost}$ \cite{Manas3}. Its design is inspired by the \textit{Law of Marginal Returns} from microeconomics theory, which it applies to assess the fairness in channel allocation to the radios in a wireless network. 
\textit{Cumulative X-Link-Set Weight} or CXLS$_{wt}$ metric is another TIEM, which operates at the level of \textit{X-Link-Set} (XLS), \emph{i.e.,} it takes into account the link-conflicts within a set of $X$ wireless links in a WMN and assigns every XLS a \textit{weight} (XLS$_{wt}$), which is indicative of its resilience to Interference. 
\textit{Channel Assignment Link-weight Metric} (CALM) is another reliable TIEM proposed in \cite{CALM, iCALM}. CALM is inspired by social theory and its design is based on the Durkheimian concept of a \textit{sui generis social reality}. 

\subsection{Capacity Interference Relationship : Open Questions}
A review of earlier studies that either aim to optimize network performance, or study the impact of \textit{Signal and Noise plus Interference Ratio} (SINR) on the capacity of traditional WiFi networks reveals that the relationship between the two is inverse and largely curvilinear \cite{Cur3}.

However, the true nature of this inverse relationship is unclear and no study, to the best of our knowledge, has demonstrated its statistical significance with high confidence. In the landmark work \cite{2Gupta}, authors theoretically demonstrate an inverse quadratic association between Interference and maximal achievable throughput, which we will verify in this work. 
Network Capacity and Interference share a close association, which is corroborated with substantial evidence \cite{Manas, cite6}.
However, no study has explored the nature of the relationship to ascertain if it is always curvilinear or quadratic. If not, it remains to be seen under what conditions do Interference and Network Capacity demonstrate a strong linear correlation. Further, impact of network design choices, topological constraints, and technical factors, such as network layout, type of CA scheme, PHY datarate, no of radios/node, \emph{etc.,} must be investigated and analyzed to determine if the relationship remains unaltered due to these modifications.

Another important aspect is the statistical credibility of the discovered relationship \emph{i.e.,} can the variation in one be explained through the other? By varying the intensity of Interference, to what degree can the consequent variation in Network Capacity be explained by the observed relationship, and vice-versa. 
Further, TIEMs are used to represent the impact of Interference on network performance, especially Network Capacity. A substantial amount of  research literature is based on the premise that an estimate showing lower levels of Interference in a WMN implies high Network Capacity. Evaluation of TIEMs is limited to observing an inverse pattern between their recorded values and the recorded Network Capacity. 
Again, the exact nature of association between TIEMs and Network Capacity has not been investigated. This creates a problem which we call, \textit{the double unknown}, \emph{i.e.,} \ws{neither} the relationship between 
Interference \& Network Capacity, nor TIEMs \& Network Capacity \ws{has been} 
precisely identified, yet we try to compare various TIEMs with each other employing Network Capacity as the network parameter of choice.
A deeper insight into the association between them will provide better benchmarks for evaluating TIEMs and their ability to model Interference efficiently. 

\wx{These questions hold great relevance in \ws{the} context of the 
\ws{IEEE}802.11ax standard which is designed to cater to the needs of dense and ultra-dense
WiFi deployments \cite{ax}. As discussed earlier, densification of OBSS and multi-AP deployments will require novel solutions to the age old problem of Interference mitigation, avoidance, and cancellation.} The impetus on the self-organizing EasyMesh technology by Wi-Fi Alliance \cite{easy} also adds to the importance of the questions raised and addressed in this work. 

\subsection{Research Contributions}
In this work, we investigate the \ws{relationship} 
of Interference and Network Capacity\ws{, and determine its nature} 
by subjecting the observed results and theoretical Interference estimates to Linear and Polynomial Regression.
For 
successful regression analysis, it is imperative that \ws{the} best regression models are chosen for each scenario based on appropriate criteria. To achieve that, 
we devise a generic algorithm for Regression Model selection/rejection that is not limited to Capacity Interference Relationship (CIR)\ws{, as well as a} 
criteria for \ws{selecting} 
a suitable Alternate Regression Model. 

To generate a theoretical measure of the endemic Interference, we consider \ws{four} 
TIEMs \emph{viz.,} TID, CDAL$_{cost}$, CXLS$_{wt}$, and CALM. We also explore if a statistical correlation exists between Interference and Network Capacity, and under what conditions. Further, we determine the extent to which variation in one can be explained through the other. 
We assess the influence that WMN topology has on their relationship by considering \ws{four} 
different WMN \ws{topologies}, 
of which \ws{two} 
are planned grid WMNs \ws{while the other two topologies} 
are specifically designed to emulate real-world WMNs by considering two deployment scenarios \emph{viz.,} a \textit{sub-urban residential community} and an \textit{urban complex with open public spaces}. 

\subsection{Relevance of Findings}
Theoretical proofs of a non-linear relationship between Interference and network performance proposed in the state-of-the-art work \cite{2Gupta}, have formed the basis of the solutions to numerous optimization problems concerned with Network Capacity, node-placement, and resource allocation in conventional WiFi networks. A non-linear relationship when formulated as a constraint in an optimization problem makes it computationally resource-intensive. \wx{It also increases the convergence time, often exponentially, which is not feasible for dense and ultra-dense OBSS scenarios  and dynamic vehicular networks \cite{dense6}.} We demonstrate through extensive simulations and analysis that a non-linear relationship between the two variables does not always exist, and should not be assumed to be so. This work paves the way for an empirical and practical approach to reflect the CIR in network optimization formulations, thereby relaxing the time and computational overhead through the use of linear constraints. 

Further, several network-optimization solutions first assume a non-linear relationship, and given its resource demands, relax these constraints to arrive at simpler, more easily solved constraints. Commonly employed simplification techniques are to replace the non-linear relationship with a linear constraint, derive a less-computationally intensive heuristic, or consider simpler Interference distribution functions \cite{GuptaOpt, amaldi2008optimization}. However, these simplification techniques are seldom guided by empirical and practical considerations of the CIR. This work offers experimental evidence in support of the use of linear constraints to model their relationship, and by making the actual optimization model less resource-intensive it also reduces the need for a simpler heuristic. Thus the findings of this work will facilitate improved and quicker network optimization. \wx{This will greatly benefit the emerging dense 802.11ax HEWs, vehicular networks, and proximity-centric mobile networks.}

\section{Statistical Analysis of CIR}
Regression Analysis is a set of statistical tools that are employed to determine relationship between variables in a system \cite{Reg2}. 
These variables can be classified as \textit{dependent variables} ($D_{var}$) and \textit{independent variables} ($I_{var}$). 
Regression Analysis helps in investigating if, and how, changes in one or more $I_{vars}$ effects the $D_{var}$ being studied, and offers a regression model (RM) that explains their relationship. We analyze the CIR from both directions, \emph{i.e.,}
\begin{enumerate}
  \item $I_x T_y$ : Capacity is $D_{var}$, Interference is $I_{var}$.
 \item $T_x I_y$ : Interference is $D_{var}$, Capacity is $I_{var}$.
 \end{enumerate}
In our analysis, we choose two regression techniques, \emph{viz.,} Linear Regression and Polynomial Regression. It is noteworthy, that the latter is in effect a type of Multivariate Linear Regression. Our choice is predicated upon the simplicity provided by these techniques in the analysis and interpretation of the CIR. These techniques offer valuable insights into the relationship by determining its \textit{statistical significance} (P-value), and calculating the \textit{percentage of variation} in \ws{the} dependent variable (R-Squared) that can be explained by the change in \ws{the} independent variable. In non-linear regression, P-value and R-Squared are not feasible, which makes it complicated to use. Further, Linear Regression facilitates determination of scenarios in which Network Capacity and Interference may have a linear relationship, and up to what extent, based on the observed Correlation Coefficient (CC). As Capacity and Interference are expected to have a non-linear relationship \cite{2Gupta}, Polynomial Regression fits non-linear data along a curve by expressing a dependent variable ($Y$) as an $n^{th}$ degree polynomial of one or more independent variables, where \ws{$n>1$}. 
\ws{W}e evaluate\ws{d numerous} 
regression models, and for efficient categorization we classify the relationship between Interference and Capacity based on the following \emph{P-value} criteria.
\begin{enumerate}
  \item \emph{P-value} $<$ $0.001$ : \textit{Highly Statistically Significant} (HSS).
  \item  $0.001$ \ws{$\leq$} 
  \emph{P-value} $<$ $0.05$   : \textit{Statistically Significant} (SS).
  \item \emph{P-value} \ws{$\geq$} 
  $0.05$  : \textit{Not Statistically Significant} (NSS).
\end{enumerate}
Further, we also consider a \textit{level of risk} ($\alpha = 0.05$), in accepting that a relationship between Capacity and Interference exists, when actually it does not.
We also account for the \textit{Outliers}, which are aberrations that usually have a disproportionate influence on statistical analysis, and can lead to misleading interpretations and conclusions.

\section{Selection of Regression Models}
The CIR for a given scenario can be explained through several regression models, which makes selection of the right regression model a crucial task. The right regression model not only captures the relationship with high statistical accuracy, it also gives an insight into the relationship in terms of its statistical significance, explanation of the variation in the response variable, \emph{etc.} However, often the question arises as to which regression model is to be chosen when R-squared values of two models are comparable. Clearly, selecting a regression model requires some objective criteria. It also involves some subjectivity, especially in terms of the type of regression analysis we choose to carry out, based on the expected nature of relationship. 
We propose a generic \textit{Selection Algorithm for Regression Model} (SAM) for CIR in Algorithm~\ref{select}. It is generic as it does not depend upon the WMN topology, \ws{physical layer (PHY)} data-rate, IEEE802.11 standard or any other WMN design choice. It only concerns itself with statistical aspects of the association between Capacity and Interference. 
SAM considers the universal set of all regression models that are run, and outputs two models that best explain the CIR for the given scenario \emph{viz.,} the Best Regression Model (BRM), and the Alternate Regression Model (ARM). It begins by pruning the regression models that are not statistically significant. For polynomial regression models of $n^{th}$ degree, SAM rejects the model if the $n^{th}$ term is not statistically significant, else the model is considered for further analysis regardless of the statistical significance of lower-order terms. We introduce an element of \textit{empirical propriety} by placing an upper-limit on the number of outlier data points in the model, which is denoted by $\Omega$ \ws{and defined by the user.} 
\renewcommand{\algorithmicrequire}{\textbf{Input:}}
\renewcommand{\algorithmicensure}{\textbf{Output:}}
\begin{algorithm}
\caption{Selection Algorithm for Regression Model}
\label{select}
\begin{algorithmic}[1]
{\fontsize{9}{10}
\REQUIRE $U_{RM}$; $RM_i \in U_{RM}$, $i \in \{1 \ldots n\}$; $S_{RM}$; $\alpha$; $\Omega$; $Out_i$ $\forall$ $RM_i$; $\rho_i$ $\forall$ $RM_i$;  $_{A}R_i^2$ $\forall$ $RM_i$; $P_i$ $\forall$ $Polynomial$ $RM_i$ \\
\ENSURE \textit{BRM, ARM} \\
\textbf{Notations} $:$ $U_{RM}$ $\leftarrow$  Set of all Regression Models RM$_i$ considered for the relationship; $n$ $\leftarrow$ $\mod{(U_{RM})}$; $S_{RM}$ $\leftarrow$ Set of statistically significant Regression Models; $\alpha$ $\leftarrow$ Significance Level;  $\Omega$ $\leftarrow$ Outlier Threshold; $Out_i$ $\leftarrow$ Number of Outliers observed in $RM_i$; $\rho_i$ $\leftarrow$ \textit{P-value} for $RM_i$ ; $_{A}R_i^2$ $\leftarrow$ Adjusted \textit{R-squared} value for $RM_i$; $P_i$ $\leftarrow$ \textit{P value} for higher-order polynomial term in $RM_i$; $BRM$ $\leftarrow$ Best Regression Model; $ARM$ $\leftarrow$ Alternate (Second Best) Regression Model. \\\vspace*{-1.5ex}
\hspace*{-15pt}\line(1,0){251}
\STATE $_{A}R_{max_1}^2$, $_{A}R_{max_2}^2$ $\leftarrow 0$ \COMMENT {$_{A}R_{max_1}^2$ $\&$ $_{A}R_{max_2}^2$  $\leftarrow$ Largest and Second-largest Adjusted R-squared ($_{A}R_i^2$) values, respectively.}
\FOR {$i \in \{1 \ldots n\}$}
\IF {$((\rho_i > \alpha)$ $\|$ $((RM_i \leftarrow Polynomial Model)$ $\&\&$ $(P_i > \alpha))$ $\|$ $( Out_i > \Omega))$}  
\STATE Reject $RM_i$
\ELSE 
\STATE $S_{PRM}$ $\leftarrow$ $S_{RM} \cup RM_i $
\STATE $_{A}R_{max_1}^2$, $_{A}R_{max_2}^2$ $\leftarrow GetMax(_{A}R_i^2)$
\ENDIF
\ENDFOR
\STATE $m \leftarrow \mod{(S_{RM})}$
\FOR {$j \in \{1 \ldots m\}$}
\IF {($_{A}R_{j}^2$ == $_{A}R_{max_1}^2$)}
\STATE $BRM$ $\leftarrow$ $RM_j$ 
\ELSIF {($_{A}R_{j}^2$ == $_{A}R_{max_2}^2$)}
\STATE $ARM$ $\leftarrow$ $RM_j$ 
\ENDIF
\ENDFOR
}
\end{algorithmic}
\end{algorithm}
In our analysis, we consider $\Omega$ as \textquotedblleft 10\% of the total number of observations\textquotedblright. Finally, SAM uses Adjusted R-squared ($_{A}R^2$) to assess every regression model\ws{'}s explanatory ability and considers two models with the highest $_{A}R^2$ values \cite{Reg2}.
Thus, SAM relies on several criteria for model selection and not just the ability of the model to explain the variation in the response variable. So, even if an $n^{th}$ degree polynomial model boasts of a higher $_{A}R^2$ value, it is rejected if its $n^{th}$ term is not significant. Likewise, regardless of impressive $_{A}R^2$ values, or high significance of a model, if the number of Outliers breaches the acceptable threshold, the model is rejected. These factors make SAM a robust and comprehensive regression model selection mechanism.\\\vspace*{-12pt}
In the next section, we discuss various aspects of WMN design considered in the ns-3 simulations.

\section{WMN Design Considerations}
\subsection{Factors Affecting the Choice of WMN Topology}
Topology of WiFi deployments shapes the Interference dynamics of wireless transmissions, and it is for this reason that topology control is a popular tool of Interference alleviation in wireless networks \cite{Topo1}. 
We classify the conventional WMN layouts into three broad categories, \emph{viz.,} \textit{Grid WMN (GWMN)}, \textit{Random WMN (RWMN)}, and \textit{Planned Real-world WMN (PWMN)}. GWMNs \ws{perform} 
better than RWMNs in terms of network coverage area, and offer a wider network span than the latter \cite{Grid2}.
PWMNs are real-world communication networks which when theoretically modeled, result in complex graphs that conform neither to grid nor to arbitrary WMN layouts \cite{Complex}. 
\begin{figure*}[ht!]
  \centering%
  \begin{tabular}{cc}
   \scalebox{0.9}{\subfloat[GWMN$_{5\times5}$]{\includegraphics[width=.195\linewidth]{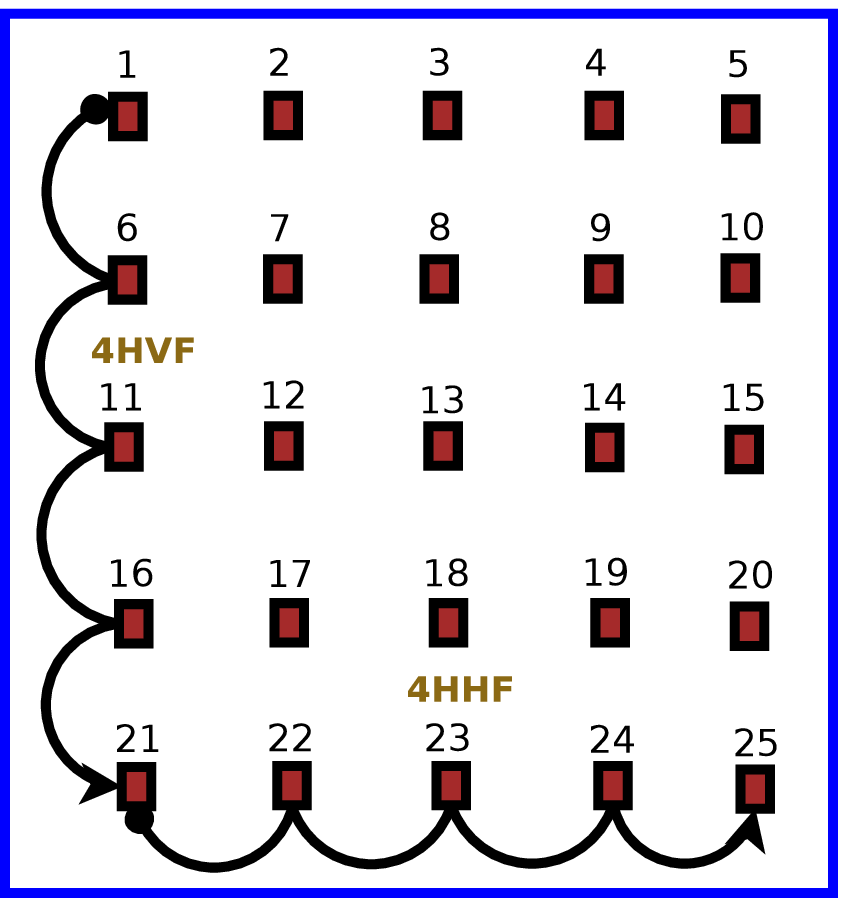}}\hspace{3em}%
   	    	\subfloat[GWMN$_{7\times7}$] {\includegraphics[width=.20\linewidth]{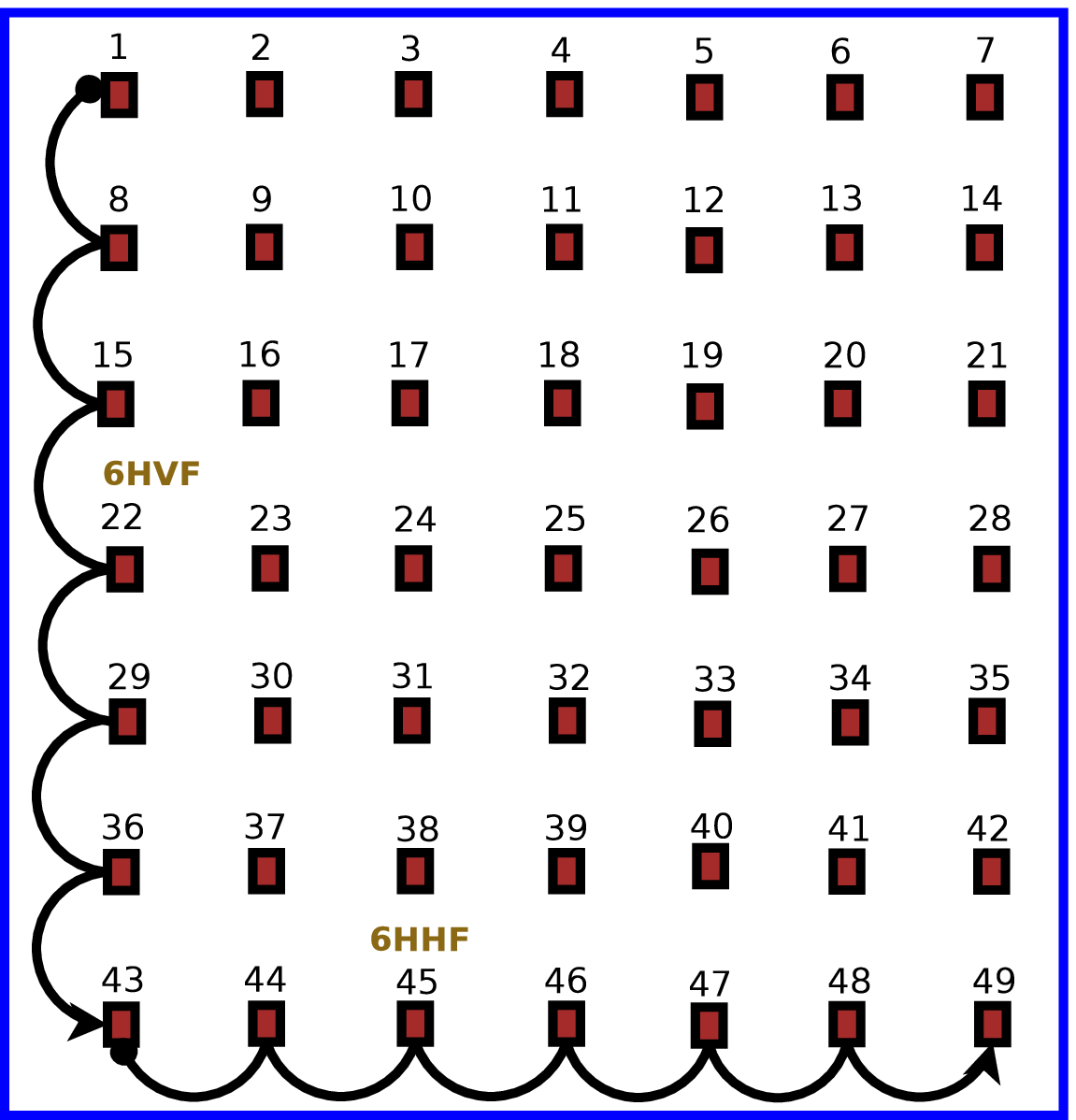}}\hspace{3em}}%
	\scalebox{0.9}{\subfloat[PWMN$_{25}$] {\includegraphics[width=.25\linewidth]{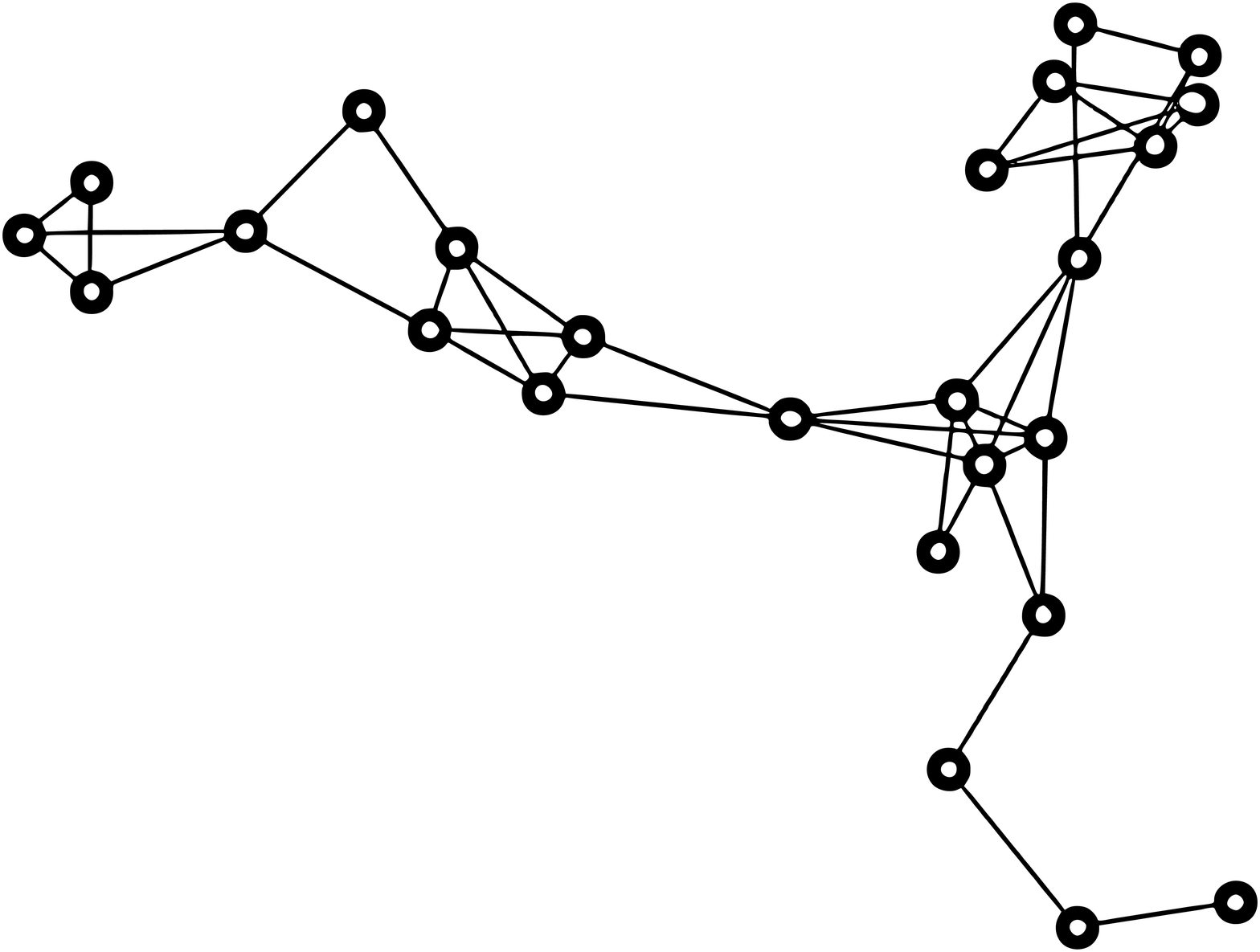}}\hspace{3em}%
		 \subfloat[PWMN$_{50}$]{\includegraphics[width=.25\linewidth]{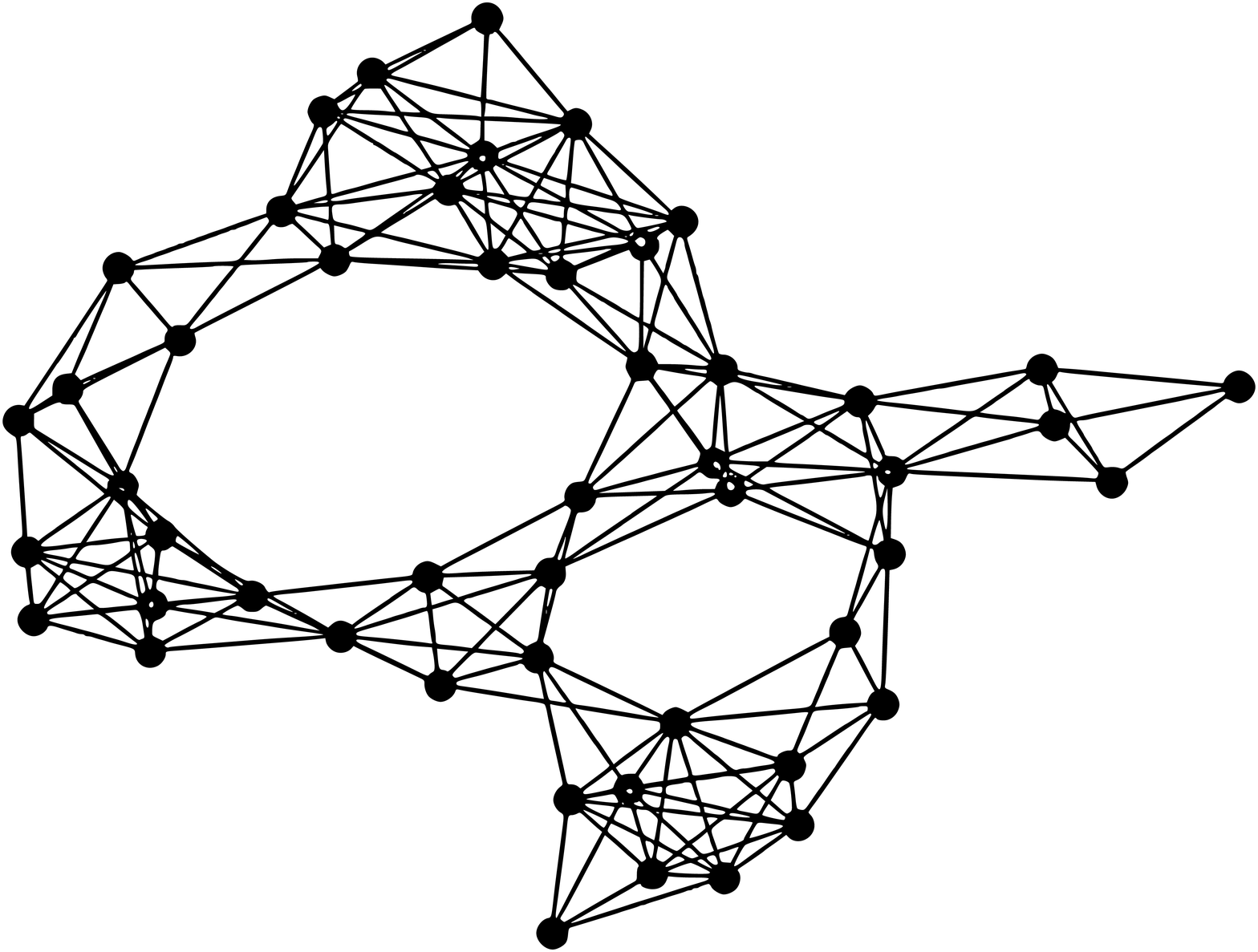}}}
    \end{tabular}
    \caption{WMN Topologies Considered for Simulation.} 
     \label{wmntopo}
\end{figure*}

\begin{table*} [h!]
\small
\centering
\begin{tabular}{|P{1.5cm}|P{3.5cm}|P{1.4cm}|P{1.4cm}|P{1.7cm}|P{1.7cm}|}
\hline
\multicolumn{1}{|c|}{\textbf{Parameter}}&\textbf{Real-World Networks}&\textbf{PWMN$_{25}$}&\textbf{PWMN$_{50}$}&\textbf{GWMN$_{5\times5}$}&\textbf{GWMN$_{7\times7}$}\\
\hline  
$\delta$&$0.05 - 0.1$&0.067&0.073&0.67&0.036\\
\hline 
$\varepsilon_{min}$ (m)&$2 - 22$&14.86&7.07&250&250\\
\hline  
$T$&$0.1 - 0.8$&0.29&0.37&NA&NA\\
\hline 
\end{tabular}
\caption{Global Parameters' Values for Simulated WMNs.}
\label{global1}
\end{table*}
Thus, while GWMNs are the preferred choice in scientific studies and industrial applications, PWMNs constitute the class of real-world WMNs which have high relevance from both, technological and socio-economic perspectives \cite{PWMN1}. It is then imperative 
that the CIR is investigated in both GWMNs and PWMNs.

\subsection{WMN Topologies Simulated in ns-3}
Bearing these facts in mind, we consider four conventional WMN topologies for the simulations which includes two GWMNs and two PWMNs. We consider GWMNs of two square-grid topologies, $5\times5$ and $7\times7$, labeled GWMN$_{5\times5}$ and GWMN$_{7\times7}$, and depicted in Figure~\ref{wmntopo}~(a) and Figure~\ref{wmntopo}~(b), respectively. Further, we design two PWMNs consisting of 25 nodes and 50 nodes, spanning a simulated environment of $1000m\times1000m$ and $1500m\times1500m$, respectively. They are referred to as PWMN$_{25}$ and PWMN$_{50}$, and presented in Figure~\ref{wmntopo}~(c), and Figure~\ref{wmntopo}~(d), respectively.

The motivation behind the design of PWMN$_{25}$ is to simulate a \textit{suburban row-housing complex}, with linear arrangement of houses (end-user nodes) located along three pathways/streets creating a relatively sparse network. Through PWMN$_{50}$, we aim to create some semblance of an \textit{urban landscape} that includes clusters of residential and office structures, situated on the periphery of two \textit{open public spaces}. \wx{Our topological choice is also guided by the premise that the upcoming IEEE802.11ax amendment will lead to a densification of existing WiFi deployments, and the CIR analysis of present networks will aid in ensuring optimal AP placement and inter-AP distance in dense OBSS scenarios.}
In Table~\ref{global1}, we present the values of three global parameters for real-world networks \emph{viz.,} Network Density ($\delta$), Radius ($\varepsilon_{min}$), and Transitivity Coefficient ($T$). 
Values of all three parameters for both planned WMNs lie within the expected range, which validates our contention that the twin PWMNs emulate real-world WMNs. In contrast, GWMNs do not demonstrate the topological characteristics of real-world deployments. 
\renewcommand{\algorithmicrequire}{\textbf{Input:}}
\renewcommand{\algorithmicensure}{\textbf{Output:}}
\begin{algorithm}[htb!] 
\caption{Interference-aware CA Generator.}
\label{ICAG}
\begin{algorithmic}[1]
{\fontsize{9}{10}
\REQUIRE $G = (V,E)$, $C =\{1 \dots n\}$, $STIEM$. \\
\textit{Notations} $\Rightarrow$ $G$ : WMN Graph, $Channel$ : Set of n orthogonal channels, $S_{TIEM}$ : Set of Theoretical Interference Metrics, $TIE$ : Theoretical Interference Estimate;
For $i \in V \rightarrow Adj_{i}$ : List of adjacent nodes, $ChRad_{i}$ : Channel set assigned to radios, $Seq_i$ : Node sequence number ; 
For ${i, j} \in V \rightarrow Ch_{mut}$ : Channels mutual to i \& j, $Ch_{ex}$ : Channels exclusive to either i or j.\\
\ENSURE Channel Assignment for $G$ \\\vspace*{-1.5ex}
\hspace*{-15pt}\line(1,0){251}
\STATE Select TIEM $\in S_{TIEM}$
\STATE Graph Preserving CA $\leftarrow$ $IMF (TIEM)$
   \IF {Topology Preserving CA}
\FOR {$i \in V$}
\FOR {$j \in Adj_{i}$} 
\IF {$((Seq_i < Seq_j)$  $\&\&$ $(\lvert ChRad_{i} \cap ChRad_{j}\lvert$ $==0)$ $\&\&$ $(TIE_{curr} < TIE_{prev}))$}
\STATE 	$ChRad_{j}  \leftarrow ChRad_{j} + \{Ch_{mut}\} - \{Ch_{ex}\}$ $\lvert \ \{(Ch_{mut}\in ChRad_{i})$ $\&\&$ $(Ch_{ex} \in ChRad_{j}) \}$
\ENDIF
\ENDFOR
\ENDFOR
\STATE \textit{Output Topology Preserving CA}
\ELSE
\STATE \textit{Output Graph Preserving CA}
\ENDIF

}
\end{algorithmic}
\end{algorithm}

\renewcommand{\algorithmicrequire}{\textbf{Input:}}
\renewcommand{\algorithmicensure}{\textbf{Output:}}
\begin{algorithm}[htb!] 
\caption{Interference Mitigation Function.}
\label{IMF}
\begin{algorithmic}[1]
{\fontsize{9}{10}
\REQUIRE$G = (V,E)$, $C =\{1 \dots n\}$, $TIEM$, $ChRad$.\\
\textbf{Notation} $\Rightarrow$ $G$ : WMN Graph, $Channel$ : Set of n orthogonal channels, $TIEM$ : Selected Theoretical Interference Estimation Metric, $ChRad$ : Set of Channel-sets assigned to radios of all $nodes \in V$, $TIE$ : Theoretical Interference Estimate.
\ENSURE Return Minimal Interference Graph Preserving CA \\\vspace*{-1.5ex}
\hspace*{-15pt}\line(1,0){251}
\STATE $TIE_{prev} \leftarrow CalcTIE(G,ChRad,TIEM)$. \COMMENT {Initially all radios are set to default channel 1.} 
\FOR {$Node \in V$}
\FOR {$Channel \in C$}
 \STATE $Channel_{prev} \leftarrow CurrChannel(Node)$
 \IF {$Node$ is assigned $Channel$ \&\& $G$ is connected}
  \STATE $TIE_{curr} \leftarrow CalcTIE(G,ChRad,TIEM)$
  \IF {$((TIE_{curr} < TIE_{prev})$)}
  \STATE $TIE_{prev} \leftarrow TIE_{curr}$
  \ELSE 
  \STATE $Node \leftarrow Channel_{prev}$
   \ENDIF
 \ENDIF
\IF{(rand()\% 2==0)}
\STATE Return Graph Preserving CA.
\ELSE
\STATE Continue CA processing.
\ENDIF
\ENDFOR
\ENDFOR
}
\end{algorithmic}
\end{algorithm}
\subsection{Generic Interference-aware CA Generator}
  We consider a large set of a 100 CA schemes, some of which have been implemented by considering popular approaches which includes the state-of-the-art \textit{Breadth First Search CA} proposed in \cite{22Ramachandran}. A majority of the CA schemes considered in this work are generated from the \textit{Interference-aware CA Generator} (ICAG) presented in Algorithm~\ref{ICAG}. ICAG offers a generic mechanism to generate both, graph preserving CA (GPCA), and topology preserving CA (TPCA), with the help of any TIEM. The TIEM set we consider to generate CA schemes through ICAG includes $TID$, $CDAL_{cost}$, $CXLS_{wt}$, and $CALM$. ICAG invokes the \textit{Interference Mitigation Function} which starts from the most-interfering CA where all radios are assigned the default channel and generates theoretically improved Interference-aware GPCAs in each iteration. It continually lowers the value Theoretical Interference Estimate (TIE), and randomly returns GPCAs to Algorithm~\ref{ICAG}. Thereafter, if a TPCA is required, ICAG will ensure topology preservation while ensuring a minimal Interference estimate. The \textit{forward correction algorithm} for topology preservation is similar to the one proposed in \cite{Manas2}.
  
\section{Experimental Set-up}
\wx{Multi-hop data transmissions are the characteristic feature of WMNs, and with upcoming dense multi-AP deployments the number of hops are expected to increase significantly. To create a maximal Interference scenario anticipated in dense WiFi deployments, we design test cases in which all nodes of the network participate in communication of data traffic.}
A 10 MB datafile is sent across every source-destination pair via multi-hop transmissions or $n$-Hop-Flows (nHFs). 
We create a diverse traffic scenario which includes source-destination pairs that are over 10 hops away, and others which are neighbors that communicate directly. Thus, both 1HFs and 10HFs are active in the simulated environment. 
\begin{table} [t]
\vspace{0.1cm}
\caption{ns-3 Simulation Parameters.}\vspace*{-1ex}
\small
\center 
\begin{tabular}{|l|p{3.2cm}|}
\hline
\bfseries
 Parameter&\bfseries Value \\ [0.2ex]
 \hline
\hline
IEEE Standard & \mbox{802.11g (2.4 GHz)$^\dag$ \&}  \mbox{802.11n (5 GHz)$^\P$}\\
\hline
No. of Radios/Node& \mbox{2$^\dag$, 3$^\P$}   \\
\hline
Range of Radios&250m \\ 
\hline
Orthogonal Channels&\mbox{3$^\dag$, 4$^\P$}  \\
\hline
802.11g/n PHY \ws{data rate} &\mbox{9 Mbps$^\dag$ \& 54 Mbps$^\P$}  \\
\hline
File size &10 MB  \\
\hline
Maximum Segment Size (TCP)&1 KB   \\
\hline
MAC Fragmentation Threshold&2200 Bytes  \\
\hline
RTS/CTS &Enabled  \\
\hline
Routing Protocol &OLSR    \\
\hline
Propagation Delay Model&Constant Speed \\
\hline
Propagation Loss Model&Range Propagation\\
\hline
Transmission Power&16 dBm\\
\hline
\end{tabular}
\label{sim}
\begin{threeparttable}
\footnotesize
\begin{tablenotes}
        \mbox{\item[$^\dag$] GWMN ;\item[$^\P$] PWMN }
    \end{tablenotes}
     \end{threeparttable}
\end{table}
Experiments are carried out in \mbox{ns-3} and the simulation parameters are presented in Table~\ref{sim}. We make use of the native \mbox{ns-3} \textit{BulkSendApplication} to determine the Network Capacity by observing the \textit{Network Aggregate Throughput} (NAT) in Mbps, for every CA scheme. Our aim is to determine and analyze the CIR under varying topological configurations. Thus, we create a set of 48 test-scenarios (TS) by varying the WMN topology, the CA type, and the TIEM used. They are labeled as TS$_{i}$, where $i \in \{1\ldots 48\}$. In TS$_{1}$ to TS$_{4}$, we consider generic CA type in GWMN$_{5\times5}$, considering one of the four TIEMs as a response variable in each TS. Further, for every combination of WMN layout, CA type, and TIEM, we approach the CIR from both directions \emph{viz.,} $I_x T_y$ and $T_x I_y$, explained earlier. Therefore in TS$_{4}$ to TS$_{8}$, we consider the TIEMs as predictors and NAT as the response variable. A similar pattern is followed for the other three WMN layouts as well. Further, we also consider two other CA types, \emph{viz.,} GPCAs and TPCAs, keeping the WMN topology constant as GWMN$_{5\times5}$, and investigate the relationship between TIEMs and Capacity in both configurations, \emph{i.e.,} $I_x T_y$ and $T_x I_y$.

\begin{figure*}[ht!]
  \centering%
  \begin{tabular}{cc}
   \subfloat[SS versus NSS : BRMs]{\includegraphics[width=.33\linewidth]{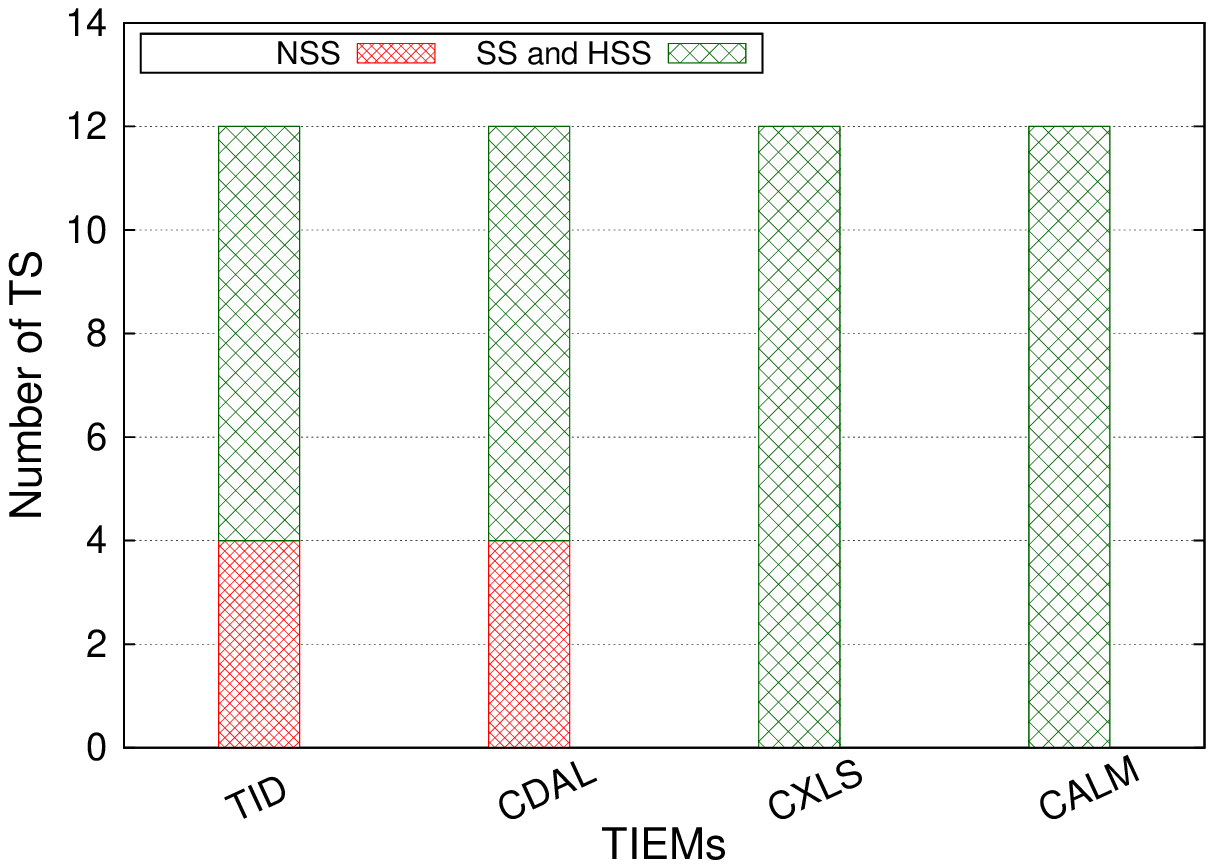}}\hfill%
    \subfloat[SS versus NSS : ARMs] {\includegraphics[width=.33\linewidth]{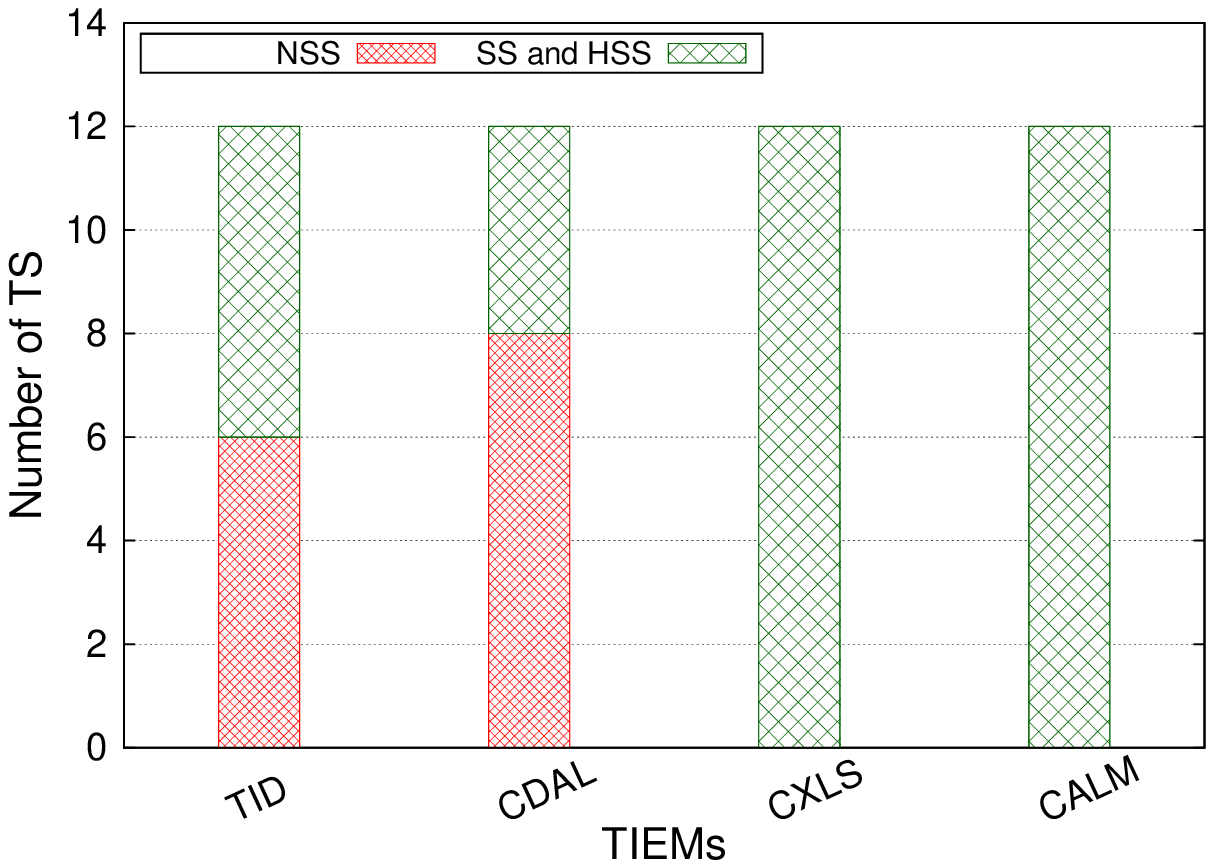}}\hfill%
	\subfloat[HSS versus SS : BRMs]{\includegraphics[width=.33\linewidth]{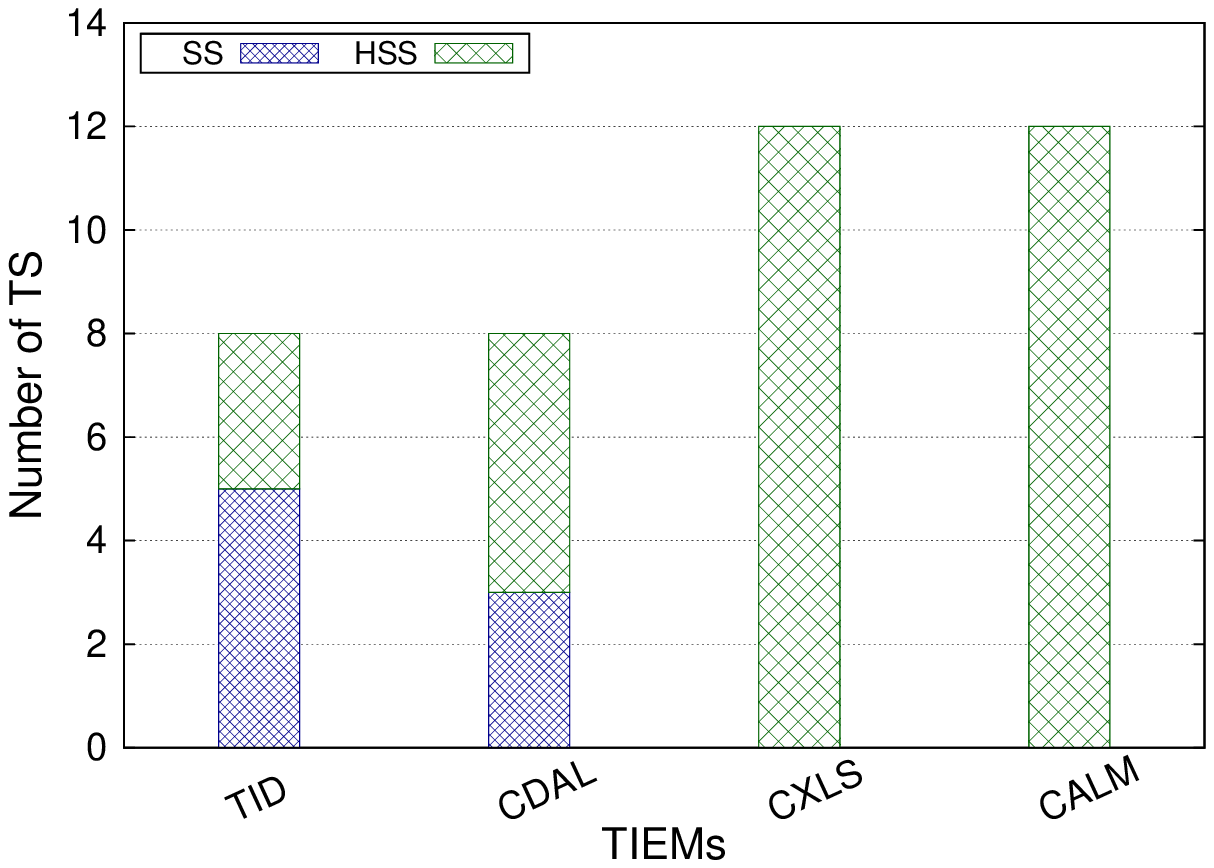}}\\
	\subfloat[HSS versus SS : ARMs.] {\includegraphics[width=.33\linewidth]{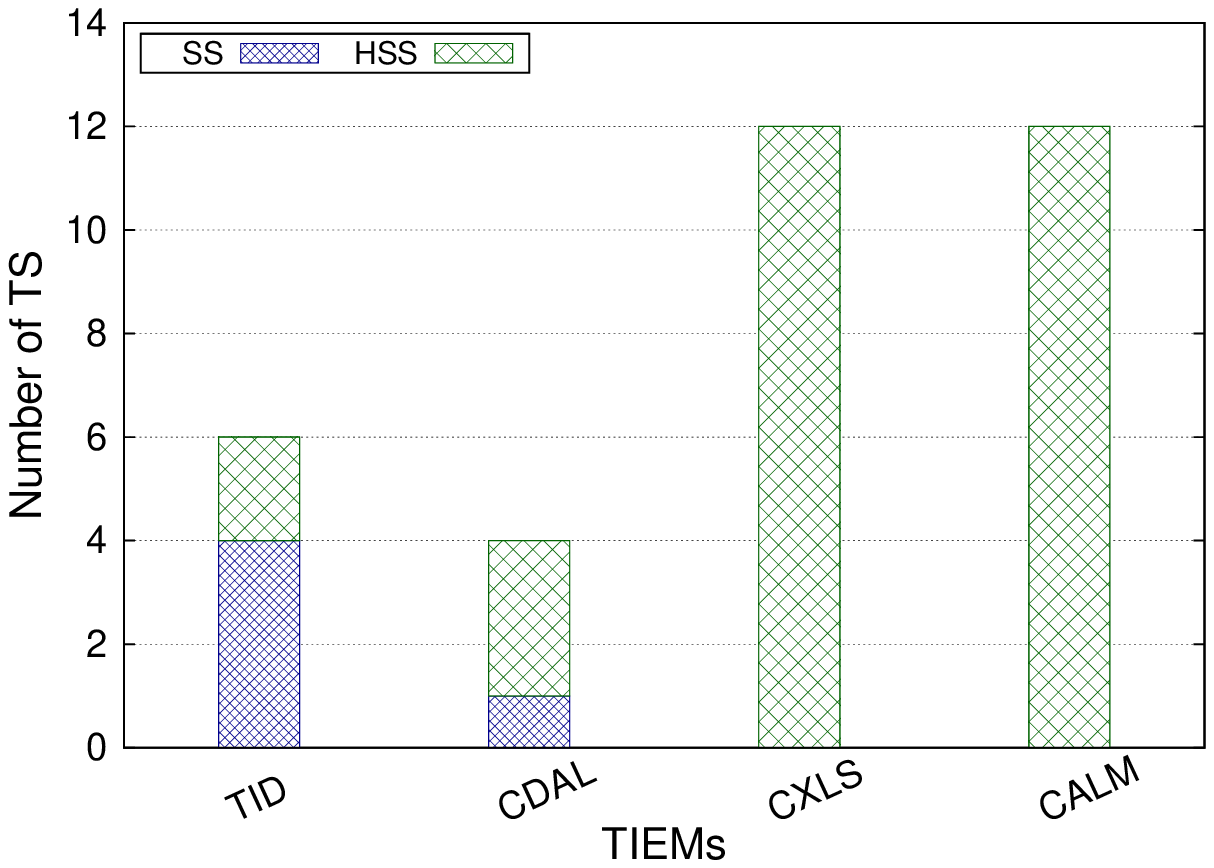}}\hfill%
    \subfloat[Linear versus Quadratic : TIEMs.] {\includegraphics[width=.33\linewidth]{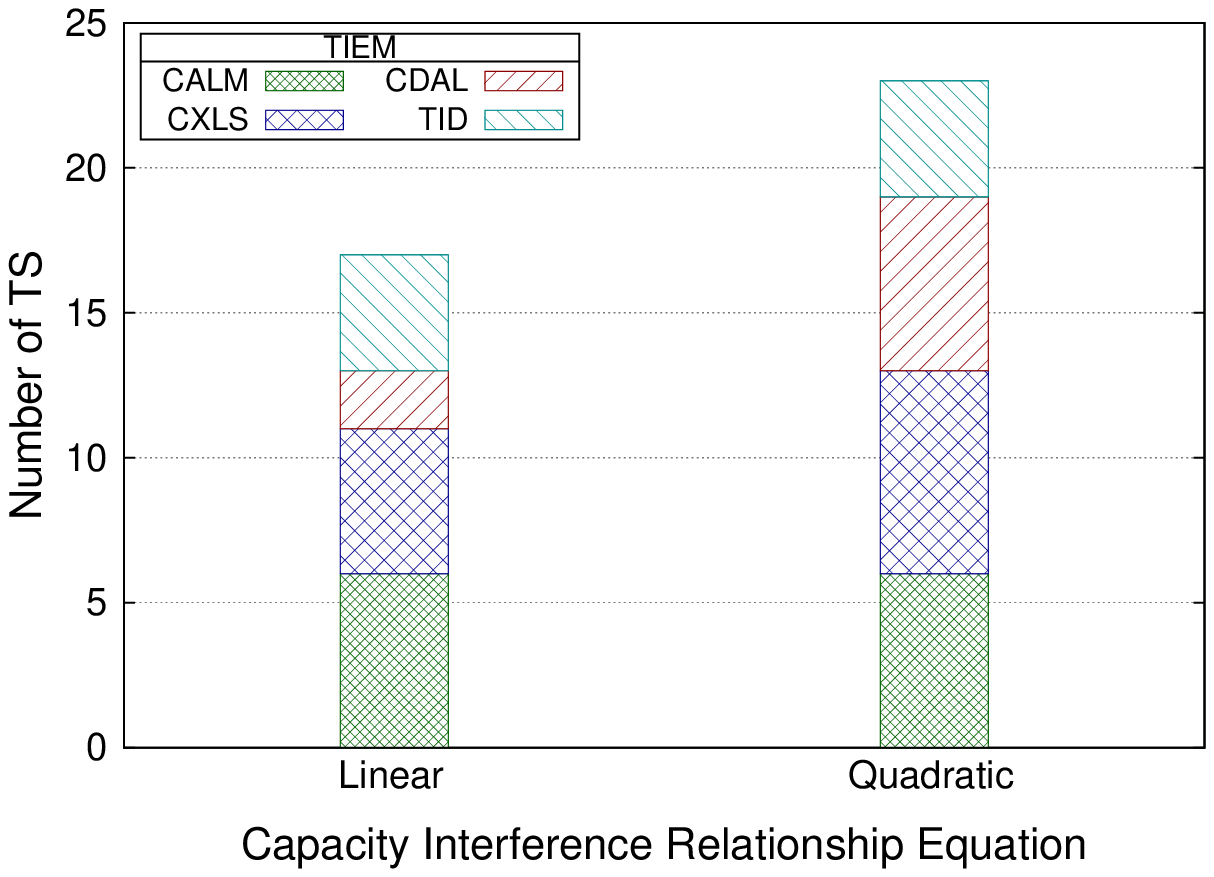}}\hfill
    \subfloat[WMN Topology and R-sq.]{\includegraphics[width=.33\linewidth]{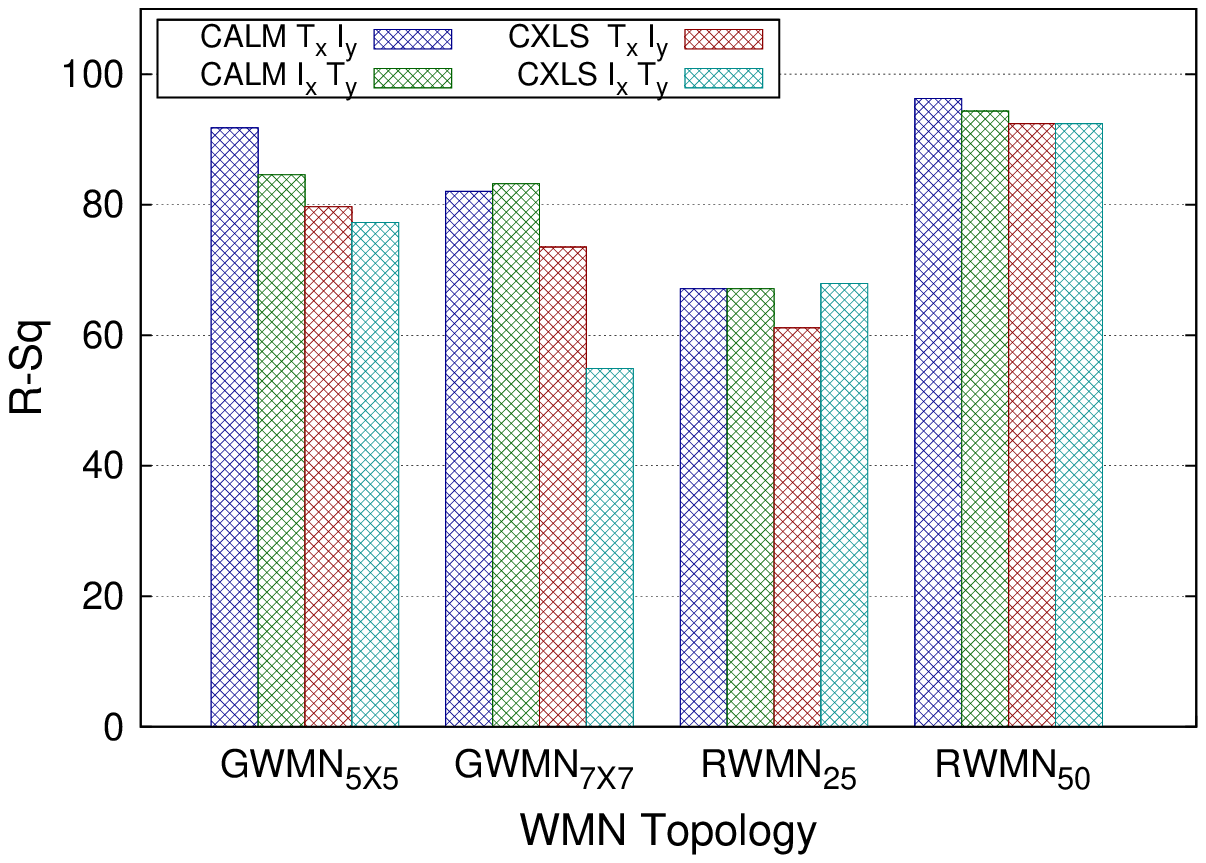}}
	\end{tabular}
     \vspace*{0.1cm}
   \caption{Regression Analysis of the Capacity Interference Relationship.} 
     \label{ANA1}
     \vspace*{-0.3cm}
\end{figure*}

\section{Results and Analysis}

For every CA scheme implemented on the four conventional WMN topologies, the simulations are run on ns-3 and NAT values are observed. Thereafter, theoretical estimates of Interference are generated for each CA, through the four TIEMs \emph{viz.,} CALM, CDAL$_{cost}$, CXLS$_{wt}$, and TID. 
For each TS$_{i}$, after pruning through several undesirable regression models, we select the \textit{Best Regression Model} (BRM) and the \textit{Alternate Regression Model} (ARM) through the SAM algorithm. Several aspects of the CIR are expressed by the regression model through the parameters discussed earlier such as R-sq, Correlation Coefficient, Outliers, etc. We analyze the raw data collected for each of these parameters and present concrete observations across four dimensions in the following sub-sections. 

\subsection{Statistical Significance of the CIR}
In 40 out of the 48 test-scenarios, a \textit{statistically significant} or a \textit{highly statistically significant} relationship is present. If we also consider the ARMs, for 6 additional scenarios the relationship is NSS, \emph{i.e.,} only one SS/HSS regression model exists for each of these 6 scenarios. The results are illustrated in Figure~\ref{ANA1}~(a) and Figure~\ref{ANA1}~(b) for BRMs \& ARMs, respectively. It can be discerned that with CALM and CXLS as the TIEMs, the CIR is at least statistically significant for all scenarios. In contrast, in all scenarios where the relationship is considered non-significant by the RM, the TIEMs used are CDAL and TID. 
Further analysis of statistical significance of BRMs \& ARMs presented in Figure~\ref{ANA1}~(c) and Figure~\ref{ANA1}~(d), respectively shows that only RMs with CALM and CXLS as TIEMs exhibit high statistical significance, while RMs involving CDAL and TID as TIEMs are just statistically significant.
\subsection{The Nature of CIR}

It can be inferred from the analysis that Capacity and Interference have a statistically significant relationship which conforms to the theoretical hypothesis that they share an inverse relationship. The nature of this relationship is considered to be curvilinear in general and is shown to be quadratic in earlier state-of-the-art works  \cite{2Gupta, Capacity}. 
A remarkable finding of our work is that the nature of CIR is not always non-linear. Although CIR remains to be quadratic in over half the scenarios, we encounter a linear correlation between Network Capacity and Interference estimates in 47\% cases.
A deeper analysis of with respect to TIEMs is presented in Figure~\ref{ANA1}~(e). Apart from CDAL, where the number of linear models is half that of quadratic, other TIEMs generate an equal number of linear and quadratic models. 
The linear nature of CIR does not appear to be an outcome of specific conditions, and presents itself consistently independent of the TIEM and WMN layout. Thus, the primary and a rather counter-intuitive finding of this work is that the CIR can be linear in certain scenarios. \wx{This inference will hold great relevance in dense OBSS scenarios, especially with respect to optimal AP-placement and Network Capacity optimization in 802.11ax HEWs.}
\begin{figure}[h]
                \centering
                \includegraphics[trim=0mm -3mm 0mm 0mm, width=0.9\linewidth]{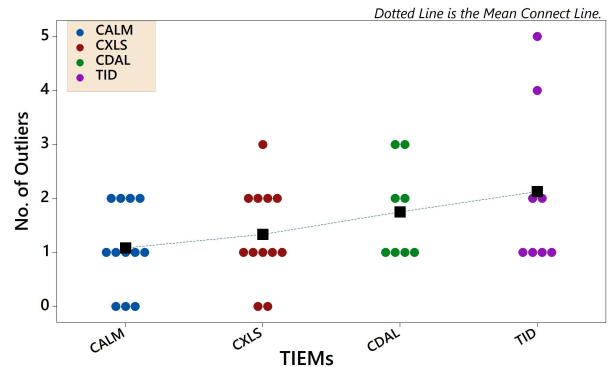}
                \caption{Outliers and TIEMs.}
                \label{OT}
%
\end{figure}

\subsection{Reliability of TIEMs}
\begin{figure}[h]
                \centering
                \includegraphics[trim=0mm -3mm 0mm 0mm, width=\linewidth]{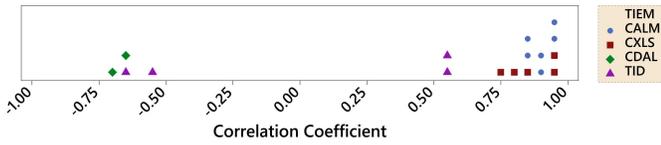}
                \vspace{0.5pt}
                \caption{Correlation Coefficient of Linear Models and TIEMs.}
                 
                \label{cc}
%
\end{figure}

Analysis of these results also sheds light on the reliability of TIEMs in effectively estimating Interference in wireless networks. Statistically significance of RMs shows that all HSS relationship scenarios either have CALM or CXLS as the TIEM while SS and NSS scenarios involve CDAL and TID as TIEMs. 
To asess how TIEMs influence the number of Outliers, we present an \textit{individual value plot} in Figure~\ref{OT}, where the mean Outlier counts are represented by black solid squares and the \textit{Mean Connect Line} joins them. It can be discerned that TIEMs clearly have a bearing upon the number of Outliers as CALM has the least mean outlier count, while TID has the maximum. The impact of TIEMs on Correlation Coefficient can be observed in the \textit{dot plot} presented in Figure~\ref{cc}. For CALM and CXLS, the CC values are close to $+1$, while for CDAL and TIEM they are relatively farther from $-1$. 
\subsection{Impact of WMN Topology on CIR}
WMN topology also seems to have an impact on the CIR, although no concrete inference can be drawn with respect to size or placement of the nodes in the WMN. We illustrate this by observing the \textit{R-sq} of scenarios involving CALM and CXLS as TIEM in all four network topologies considered.
Topological independence should ensure similar R-sq values for a single TIEM variable since the CA type is constant. However, in Figure~\ref{ANA1}~(f) a significant amount of variation can be observed for both CALM and CXLS. \wx{Clearly, with the densification of WiFi networks in the upcoming of 802.11ax implementations, placement of APs and the consequent network topology will have a great bearing upon CIR.}
\section{Conclusions and Future Work}

The foremost conclusion of this work is that the association of Network Capacity and Interference is not necessarily non-linear or quadratic, as is widely believed. A strong linear correlation exists between the two in several scenarios. 
Further, network topology influences the ability of the model to explain the change in target variable as the predictor variable changes. Finally, a regression model is only as accurate as the predictor variable used. Clearly, some TIEMs (\emph{e.g.,} CALM, CXLS) are more reliable estimates of Interference than others (\emph{e.g.,} CDAL, TID). Models involving CALM and CXLS offer high R-sq and fewer Outliers, which makes the analysis of the relationship more accurate.
\wx{As an extension to this work, we will conduct experiments on a \textit{dense WiFi network} and observe real-time SINR and Capacity values to carry out a direct evaluation. Further, we will investigate the benefits of linear correlation between Capacity and Interference in network optimization problems in dense WiFi deployments. We also intend to introduce the element of mobility in our investigations.}
\bibliography{ref_Journal}

\end{document}